# SOLAR NEUTRINO CAPTURE BY 128, 130TE ISOTOPES AND BAKSAN LARGE NEUTRINO TELESCOPE PROJECT

A. N. Fazliakhmetov[1,2*], Yu. S. Lutostansky[1], B. K. Lubsandorzhiev[2],

G. A. Koroteev[1,2], V. N. Tikhonov[1]

[1]National Research Center Kurchatov Institute, Moscow, Russia
[2]Institute for Nuclear Research, Russian Academy of Sciences, Moscow, Russia
* E-mail: fazliakhmetov@phystech.edu

This paper discusses the charge-exchange strength functions $S(E)$ of the isotopes $^{128,130}$Te. The experimental data of the $S(E)$ strength functions obtained from ($^3$He, $t$) reactions, as well as the $S(E)$ strength functions calculated in the microscopic theory of finite fermi-systems, are analyzed. The resonance structure of the strength function $S(E)$ is examined, and the Gamow-Teller and Pygmy resonances are identified. The resonance structure of the strength function $S(E)$ is vital for the calculation and analysis of the process of neutrino capture by atomic nuclei.

## 1. INTRODUCTION

The neutrino-matter interaction cross section is crucial for interpreting experimental data and designing future experiments. The value and energy dependence of the neutrino capture cross section $\sigma(E_\nu)$ depend on the structure of the charge-exchange strength function $S(E)$, which describes the intensity of transitions in the final nucleus as a function of excitation energy. The $S(E)$ strength function has a resonance nature, which is observed in almost all charge-exchange reactions [1-3] and appears mainly as a giant Gamow-Teller resonance (GTR) [4-7], Analog resonance (AR), and low-lying Pigmy resonances (PR) [8]. The resonance nature of the charge-exchange strength function $S(E)$ can greatly affect the results of theoretical calculations of neutrino capture cross sections for atomic nuclei $\sigma(E_\nu)$ [9,10], as well as matrix elements for beta-decay and double beta-decay processes.

In this paper, we analyze the experimental data on the charge-exchange reactions of $^{128}$Te($^3$He, $t$)$^{128}$I и $^{130}$Te($^3$He, $t$)$^{130}$I [11]. The interest in the isotopes $^{128}$Te and $^{130}$Te is driven by their significance in geochemical experiments [1] due to their high natural abundance of Tellurium: 31.78% for $^{128}$Te and 34.08% for $^{130}$Te. Additionally, Tellurium isotopes have a historical importance, as they were among the first isotopes for which the double beta-decay period was measured in a geochemical experiment [12, 13]. Furthermore, isotope $^{130}$Te is among the popular candidates for searching the process of neutrinoless double beta-decay [1], due to its large decay energy $Q_{\beta\beta}$=2527.51 keV [14] which is higher than most sources of natural radioactivity (for $^{128}$Te $Q_{\beta\beta}$=866.7 keV [14]), providing a larger phase space.

The SNO+ experiment plans to use a 780-ton liquid scintillator target, made of linear alkylbenzene (LAB) with dissolved Tellurium in it, with a concentration of 0.5% by mass in the first phase of the experiment to search for the $0\nu\beta\beta$ process in $^{130}$Te [15, 16]. The search for the $0\nu\beta\beta$ process in $^{130}$Te has also been conducted by the CUORE experiment, which utilizes an array of 988 TeO$_2$ crystals that serve as both a target and a cryogenic bolometric detector [17, 18, 19]. Additionally, results on the measurement of the $^{130}$Te half-life in the $2\nu\beta\beta$ and $0\nu\beta\beta$ channels have been published from the COBRA [20] and NEMO-3 [21] experiments.

The $^{130}$Te isotope is also proposed to be used to search for the $0\nu\beta\beta$ process in the Baksan Large Neutrino Telescope project. The goal of this project is to design a next-generation multipurpose neutrino detector to detect neutrino and antineutrino fluxes from the Sun, Earth, and astrophysical sources [22]. It is proposed to construct a liquid scintillation neutrino detector with a target mass of 10 kt, located in the Northern Caucasus in the underground halls of the Baksan Neutrino Observatory of the Russian Academy of Sciences, at a depth of about 4700 meters water equivalent. A prototype detector with a target mass of 0.5 t has been constructed and tested, and a prototype with a target mass of 5 t is currently under construction. The use of isotope $^{130}$Te dissolved in liquid scintillator is proposed in the third phase of the project, on a prototype with a target mass of 100 tons. Other isotopes are also being considered for use such as $^{150}$Nd, $^{115}$In.

## 2. CHARGE-EXCHANGE EXCITATIONS OF ISOTOPES $^{128}$Te AND $^{130}$Te

Fig. 1 shows a scheme of the excited isobaric states of the $^{128}$Te nucleus, which are observed in the neighbor $^{128}$I nucleus. At excitation energies above $S_n = 6826.13 \pm 5$ keV [14], there is decay into the stable nucleus $^{127}$I along with neutron emission. At lower excitation energies, transitions from the excited state of the $^{128}$I nucleus to the ground state occur first, followed by the beta-decay of $^{128}$I from the ground state to the ground state and the excited states of the $^{128}$Xe nucleus. A similar decay scheme is shown in Fig. 2 for the $^{130}$Xe nucleus: decay to $^{129}$I with neutron emission at an excitation energy greater than $S_n = 6500.33 \pm 4$ keV [14], or beta-decay from the ground state of $^{130}$I to the ground and excited states of $^{130}$Xe after all excitations with energies $E_x < S_n$ are decayed to the ground state $^{130}$I. It is important to note that because the beta-decay is to a highly excited state that decays quickly emitting accompanying γ-rays, the fraction of the decays that populate an energy region near the end point for $0\nu\beta\beta$ decay of $^{130}$I is relatively large. These γ-rays can become a background process in relevant experiments.

Experimental spectrums for the charge-exchange reactions $^{128}$Te($^3$He, $t$)$^{128}$I and $^{130}$Te($^3$He, $t$)$^{130}$I obtained with 35 keV resolution [11] are shown in Fig. 3 and Fig. 4, together with the decomposition of the spectra into individual resonances. We fitted the shape of the resonances by Breit-Wigner form, the background from the quasi-free continuum states (QFC) was approximated similarly in previous work [23]. The obtained energies are $E_{GTR}$ = 13.17 MeV, $E_{PR1}$ = 8.53 MeV, $E_{PR2}$ = 6.50 MeV for $^{128}$Te and $E_{GTR}$ = 13.84 MeV, $E_{PR1}$ = 8.21 MeV, $E_{PR2}$ = 6.66 MeV for $^{130}$Te. In the paper [11] only the rough position of the Gamow-Teller resonance $E_{GTR} \approx 14$ МэВ was given, no data on the energies of the Pygmy resonances were presented. In an earlier paper [24] spectra for charge-exchange reactions $^{128}$Te($p, n$)$^{128}$I and $^{130}$Te($p, n$)$^{130}$I were measured with 350 keV resolution and tabulated values for resonances were given: $E_{GTR}$ = 13.14 MeV for $^{128}$Te and $E_{GTR}$ = 13.59 MeV for $^{130}$Te, which is in good agreement with our fitting.

One of the main challenges in determining the strength function $S(E)$ from the experimental charge-exchange reaction spectrum is the normalization of $S(E)$. In the paper [11], the full value of $\sum B(GT)$ is not provided and the dependence of $B(GT)$ value on the excitation energy $E_x$ is only given up to 5 MeV. However, an earlier work [24] gives the values of $\sum B(GT)$ for both isotopes up to $E_x$ = 25 MeV: $\sum B(GT)$ = 40.1 or 55.7% from the maximum value $3(N-Z)$ = 72 for $^{128}$I, $\sum B(GT)$ = 45.9 or 58.8% from the maximum value $3(N-Z)$ = 78 for $^{130}$Te. The observed deficit in the sum rule for GT excitations is likely due to the *quenching*-effect [25] or a disruption in the normalization of the GT matrix elements.

3. CALCULATION OF THE STRENGTH FUNCTIONS FOR THE ISOTOPES $^{128,\ 130}$Te

The charge-exchange strength functions *S(E)* for the isotopes $^{128,\ 130}$Te were calculated within the microscopic theory of finite Fermi systems (TFFS) [26] in just the same way as this was done earlier for other nuclei in the works [10, 27, 28]. The energies of the excited states of the daughter nucleus and their matrix elements were determined by solving the set of secular TFFS equations for the effective field according to the works [26, 29]. The calculations were performed in the coordinate representation with allowance for pairing in the single-particle basis. The basis was taken in the Woods–Saxon model, and the subsequent iteration procedure was used to construct the nuclear potential. Effects of the change in the pairing gap in an external field were neglected—that is, it was assumed that $d^1_{pn} = d^2_{pn} = 0$. This is justified for an external field whose diagonal elements are zero (see book [26], p. 200).

In the present study, we used the simplified version of what was done in the paper [29] — that is, a partial agreement with the local interaction and $m^* = m$ for allowed transitions with a local nucleon–nucleon interaction $F^\omega$ in the Landau–Migdal form [26]:

$$F^\omega = C_0( f'_0 + g'_0 (\sigma_1\sigma_2)) (\tau_1\tau_2)\, \delta(r_1 - r_2), \qquad (1)$$

where $C_0 = (d\rho/d\varepsilon_F)^{-1} = 300$ MeV fm$^3$ ($\rho$ is the average nuclear-matter density), $f'_0$ and $g'_0$ are the parameters of isospin–isospin and spin–isospin quasiparticle interactions. These coupling constants are phenomenological parameters. In the present calculations, we used the values $f'_0 = 1.351 \pm 0.027$ and $g'_0 = 1.214 \pm 0.048$, which were obtained recently [30]. The energies, $E_i$, and the squares of the matrix elements, $M^2_i$, were calculated for allowed-transition-excited isobaric states of $^{128,\ 130}$I daughter nuclei. The continuous part of the spectra of the strength function *S(E)* was calculated in the same way as in the paper [9] upon taking into account Breit–Wigner broadening.

In the description of both experimental and calculated data on the strength function $S(E)$ of isotopes $^{128,\,130}$Te presented in Fig. 3 and Fig. 4. The normalization of $S(E)$ is a significant issue, we used the normalization of the TFFS equal to 81% from the maximum value $3(N-Z)$, as in the works [27, 28].

## 4. CONCLUSIONS

In this work, the resonance structures of the charge-exchange strength functions $S(E)$ of the $^{128}$Te and $^{130}$Te isotopes are studied. We have analyzed both experimental data obtained for the strength functions $S(E)$ in $(p, n)$ and $(^3$He, $t)$ reactions and the strength functions $S(E)$ calculated on the basis of the self-consistent theory of finite Fermi systems. We have studied the resonance structure of the strength function $S(E)$. The results of calculations of the energies of GTR, AR and PR resonances are in good agreement with the experimental data.

The normalization of the strength functions is of fundamental importance, especially for experimental data on charge-exchange reactions, such as $(p, n)$ and $(^3$He, $t)$, and $(\nu_e, e^-)$. In the present study, we have shown that there is an underestimation of *quenching*-effect of the sum rule both in experimental data and in the results of theoretical calculations. For the isotopes $^{128}$Te and $^{130}$Te, calculated sum $\sum B_i(\text{GT}) = \sum M_i^2$ is 81% from the maximum value $3(N-Z)$ predicted by the theory.

The resonance structure of the strength function $S(E)$ is of a great importance for the calculation and analysis of all charge-exchange processes. For us, the process of neutrino capture by atomic nuclei is of particular interest..

This work is supported in the framework of the State project ``Science'' by the Ministry of Science and Higher Education of the Russian Federation under the contract 075-15-2024-541.

FIGURE CAPTIONS

Fig. 1. Scheme of charge-exchange excitations of the $^{128}$Te nucleus in the $^{128}$Te($^3$He, *t*)$^{128}$I reaction, with decay into the stable $^{127}$I with neutron emission for excitations $E_x > S_n$ or the $^{128}$I decay into $^{128}$Xe, simulating $^{128}$Te double beta-decay.

Fig. 2. Scheme of charge-exchange excitations the $^{130}$Te nucleus in the $^{130}$Te($^3$He, *t*)$^{130}$I reaction, with decay into the $^{129}$I with neutron emission for excitations $E_x > S_n$ or the $^{128}$I decay into $^{128}$Xe, simulating $^{128}$Te double beta-decay.

Fig. 3. Experimental spectrum of the charge-exchange reaction $^{128}$Te($^3$He, *t*)$^{128}$I [11] and its decomposition into individual resonances. The giant Gamow-Teller resonance (GTR) and Pigmy resonances (PR1, PR2) approximated by Breit-Wigner are identified. The background from the quasi-free continuum states (QFC) was approximated similarly in previous work [23]. The analog resonance was pre-cut from the spectrum.

Fig. 4. Experimental spectrum of the charge-exchange reaction $^{130}$Te($^3$He, *t*)$^{130}$I [11] and its decomposition into individual resonances. The giant Gamow-Teller resonance (GTR) and Pigmy resonances (PR1, PR2) approximated by Breit-Wigner are identified. The background from the quasi-free continuum states (QFC) was approximated similarly in previous work [23]. The analog resonance was pre-cut from the spectrum.

Fig. 5. The charge-exchange strength function *S(E)* of isotope $^{128}$I for GT-excitations of $^{128}$Te: 1 — experimental data [11], 2 — our TFFS calculation.

Fig. 6. The charge-exchange strength function *S(E)* of isotope $^{130}$I for GT-excitations of $^{130}$Te: 1 — experimental data [11], 2 — our TFFS calculation.

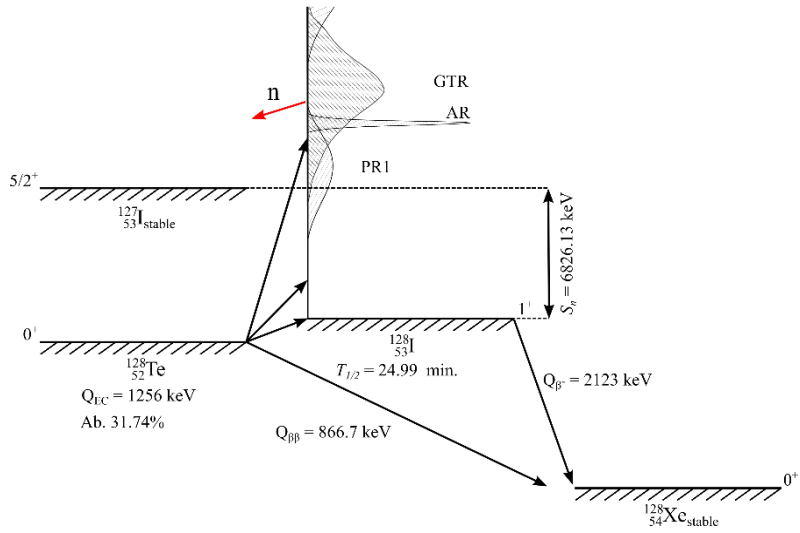

Figure 1.

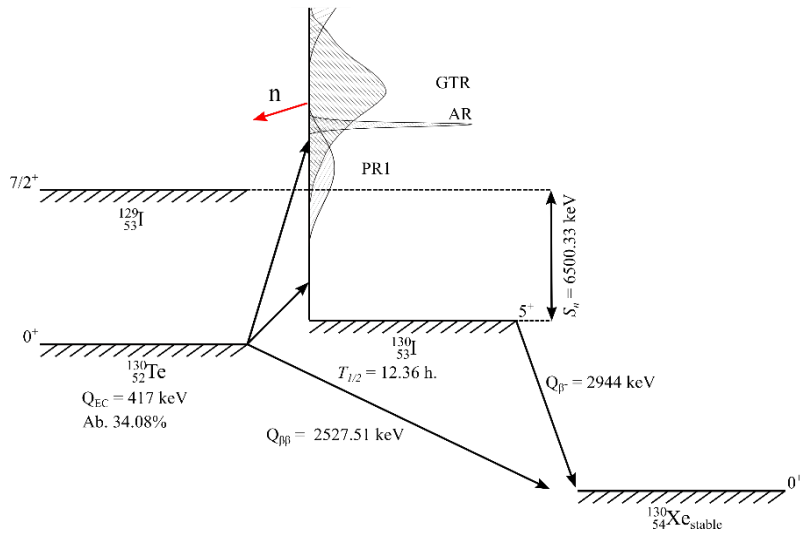

Figure 2.

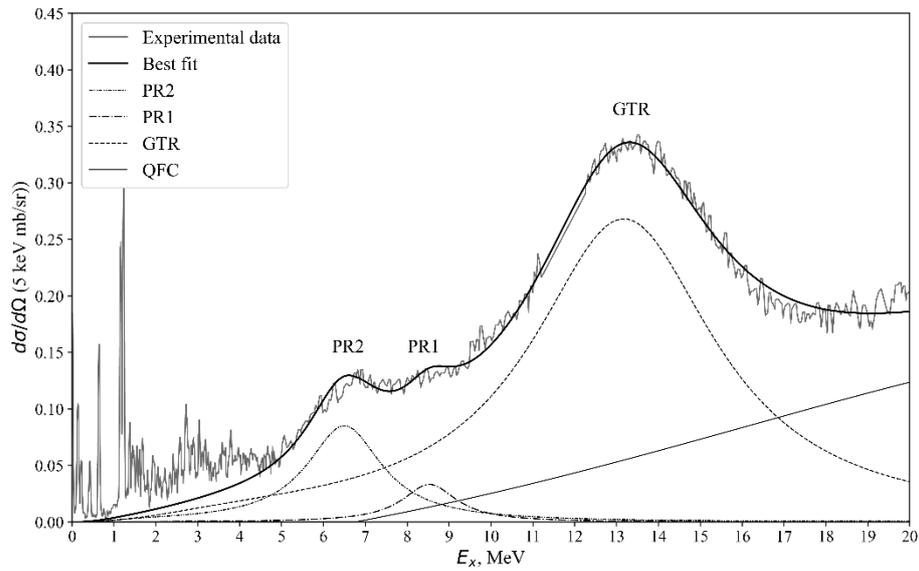

Figure 3.

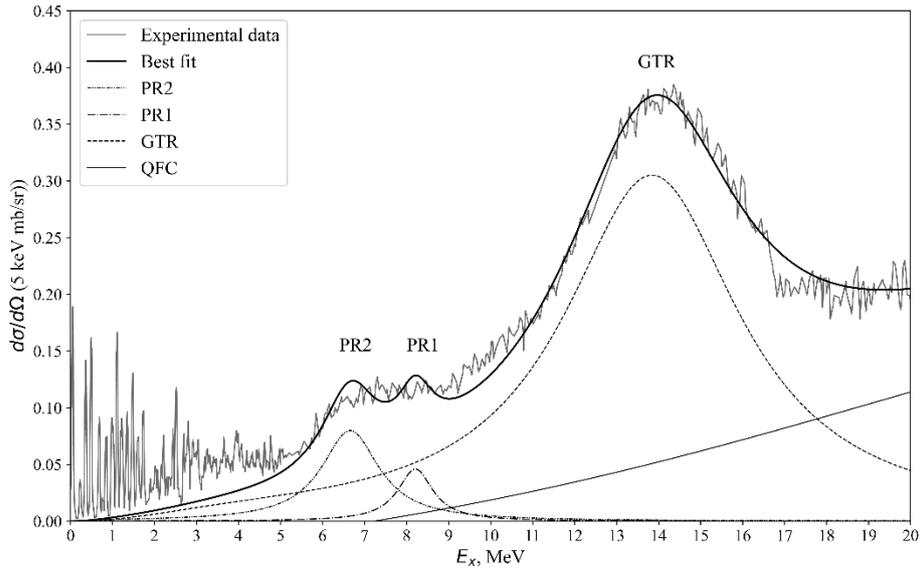

Figure 4.

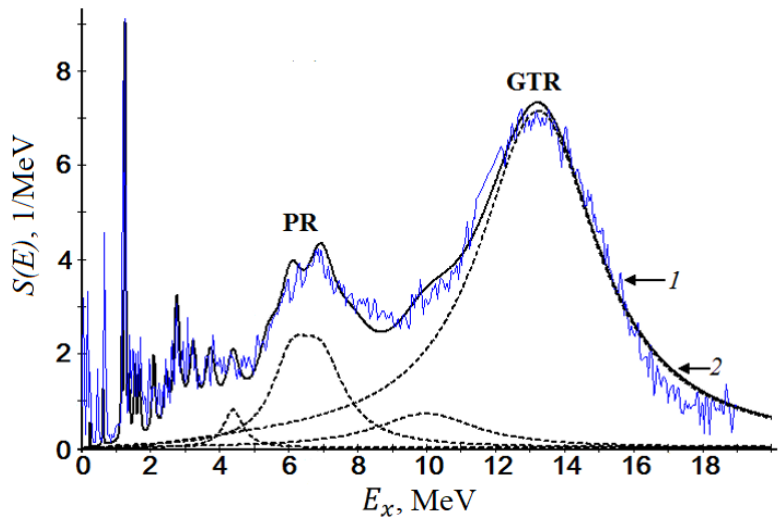

Figure 5.

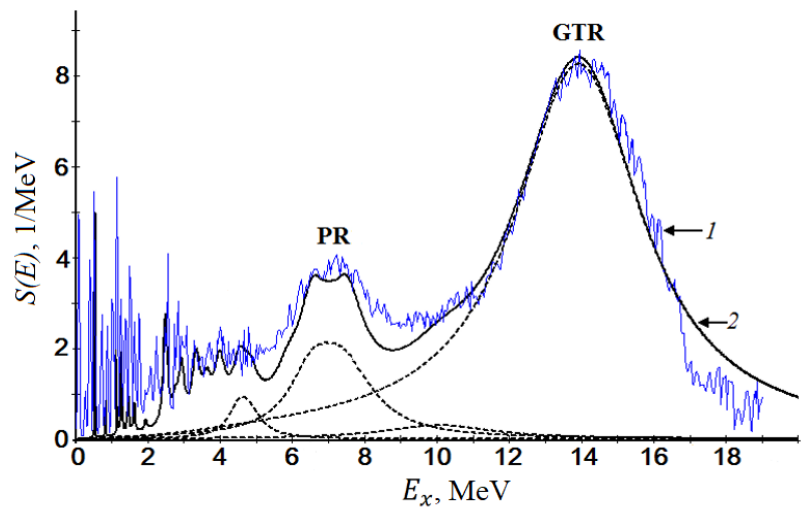

Figure 6.

10